\documentclass[12pt,preprint,apj,revtex4-1]{emulateapj}
\usepackage{amsmath}

\newcommand{\kmsMpc}{{km s$^{-1}$ Mpc$^{-1}$}~}

\newcommand{\meanmue}{{$\langle\mu\rangle_{\rm e}(r')$}~}
\newcommand{\absmue}{{$\langle\mu\rangle_{\rm e, abs} (r') $}~}
\newcommand{\Reff}{{$R_{\rm e}$}~}
\newcommand{\Reffc}{{$R_{\rm e,c}$}~}
\newcommand{\SBunit}{{mag arcsec$^{-2}$}~}

\usepackage{color}

\shorttitle{UDGs in Abell S1063 \& Abell 2744}
\shortauthors{Lee et al.}

\begin{document}

\title{
Detection of a Large Population of Ultra Diffuse Galaxies in 
Massive Galaxy Clusters: Abell S1063 and Abell 2744 
}
\author{Myung Gyoon Lee\altaffilmark{1,}, 
Jisu Kang\altaffilmark{1,}, Jeong Hwan Lee\altaffilmark{1,}, and In Sung Jang\altaffilmark{2,}}
\affil{$^1$Astronomy Program, Department of Physics and Astronomy, Seoul National University, Gwanak-gu, Seoul 08826, Korea} 
\affil{$^2$Leibniz-Institut  f{\"u}r Astrophysik Potsdam (AIP), An der
Sternwarte 16, D-14482, Potsdam, Germany}
\email{mglee@astro.snu.ac.kr} 


\begin{abstract}
We present the detection of a large population of ultra diffuse galaxies (UDGs)  in two massive galaxy clusters, Abell S1063 at $z=0.348$ and Abell 2744 at $z=0.308$,
based on F814W and F105W images in the Hubble Frontier Fields Program.
We find 47 and 40 UDGs 
in Abell S1063 and Abell 2744, respectively. 
Color-magnitude diagrams of the UDGs show that they are mostly located at the faint end of the red sequence.
From the comparison with simple stellar population models, we estimate their stellar mass to range from $10^8$ to $10^9 M_\odot$. 
Radial number density profiles of 
the UDGs show a turnover or a flattening in the central region at $r<100$ kpc.
We estimate the total masses of the UDGs using the galaxy scaling relations. 
A majority of the UDGs have total masses, $M_{200} = 10^{10}$ to $10^{11}~M_\odot$, and only a few of them
have total masses, $M_{200} = 10^{11}$ to $10^{12}~M_\odot$.   
The total number of UDGs within the virial radius is estimated to be
N(UDG)$=770\pm114$ for Abell S1063, and   
N(UDG)$=814\pm122$ for Abell 2744.
Combining these results with data in the literature, we fit the relation between
the total numbers of UDGs and the masses of their host systems for $M_{200}>10^{13} M_\odot$ with a power law,
N(UDG) $= M_{200}^{1.05\pm0.09}$. 
These results suggest that a majority of the UDGs have a dwarf galaxy origin, while only a small number of the UDGs are massive $L_*$ galaxies that failed to form a normal population of stars.
\end{abstract}


\section{INTRODUCTION}

Ultra Diffuse Galaxies (UDGs) are a mysterious type of galaxies with low surface brightness (LSB), which have larger sizes and fainter surface brightness than normal galaxies with similar luminosity. 
Since the recent revival of the UDGs in Coma  \citep{van15,kod15}, the number of the known UDGs keeps increasing.
The UDGs are found from the low-density regions to the high density environments such as galaxy clusters 
 \citep{mih15,yag16,smi16,vander16,
mar16,mer16, bel17,rom16a,rom16b}. 

The scenarios suggested to explain the origin of the UDGs can be divided roughly into two types.
First, UDGs are massive $L_{*}$ galaxies that failed to form a normal amount of stars given their dynamical mass \citep{van15,kod15,van16}.
Second, they are dwarf galaxies that were inflated due to some physical (dynamical or thermal) processes \citep{yoz15,amo16,bea16,pen16,dic17}.  
While observational evidence is being accumulated, the nature of UDGs is not yet clear and whether the UDGs are failed $L_{*}$ galaxies or inflated dwarf galaxies is still debated \citep{rom16b,zar17}.

In this study, 
we search for UDGs in two massive galaxy clusters, Abell S1063 and Abell 2744, which are 
more massive than Coma. Because they are massive clusters, it is expected that they should contain a large number of UDGs and that these UDGs may be a good sample to reveal the nature of the UDGs.  
They are part of the target galaxy clusters in the Hubble Frontier Fields (HFF) Program, for which  deep HST images are available \citep{lot16}. Abell S1063 and Abell 2744 are located at the redshift, $z=0.348$ and $z=0.308$, respectively, so their HST fields cover a relatively large fraction of each cluster. Thus these two clusters are excellent targets for the study of UDGs.
During this study, \citet{jan16} presented the discovery of UDGs 
in Abell 2744, using the HFF images.

In this study we adopt the cosmological parameters, $H_0 = 73$ \kmsMpc, $\Omega_M=0.27$, and $\Omega_\Lambda = 0.73$.
For these parameters, luminosity distance moduli of Abell S1063 
and Abell 2744 
are $(m-M)_0=41.25$ ($d= 1775$ Mpc)
and 40.94 ($d= 1540$ Mpc), 
and angular diameter distances are 978 Mpc and 901 Mpc, respectively.
Corresponding image scales of the clusters are 4.744 kpc arcsec$^{-1}$ and 4.370 kpc arcsec$^{-1}$, respectively.

The virial radius and mass of Abell S1063 are
$R_{200} = 8'.64 = 2.5$ Mpc and $M_{200}=2.7^{+0.5}_{-0.6} \times 10^{15} M_\sun$ \citep{zen16}, respectively.
Abell 2744 has a complex structure so that it is not easy to determine its mass. 
We use the virial radius and mass of the central $`a$' subcluster (called as the southern core) covered by the HFF images given by
\citet{bos06}: $R_{200} = 9'.16 = 2.4$ Mpc and $M_{200}=2.2^{+0.7}_{-0.6} \times 10^{15} M_\sun$.
Foreground reddening values toward these clusters \citep{sch11} are negligible:
$E(B-V)=0.010$ for Abell S1063 and
$E(B-V)=0.012$ for Abell 2744.

\section{DATA AND DATA REDUCTION}

\subsection{Data}

\begin{figure*}
	\centering
	\includegraphics[scale=0.9]{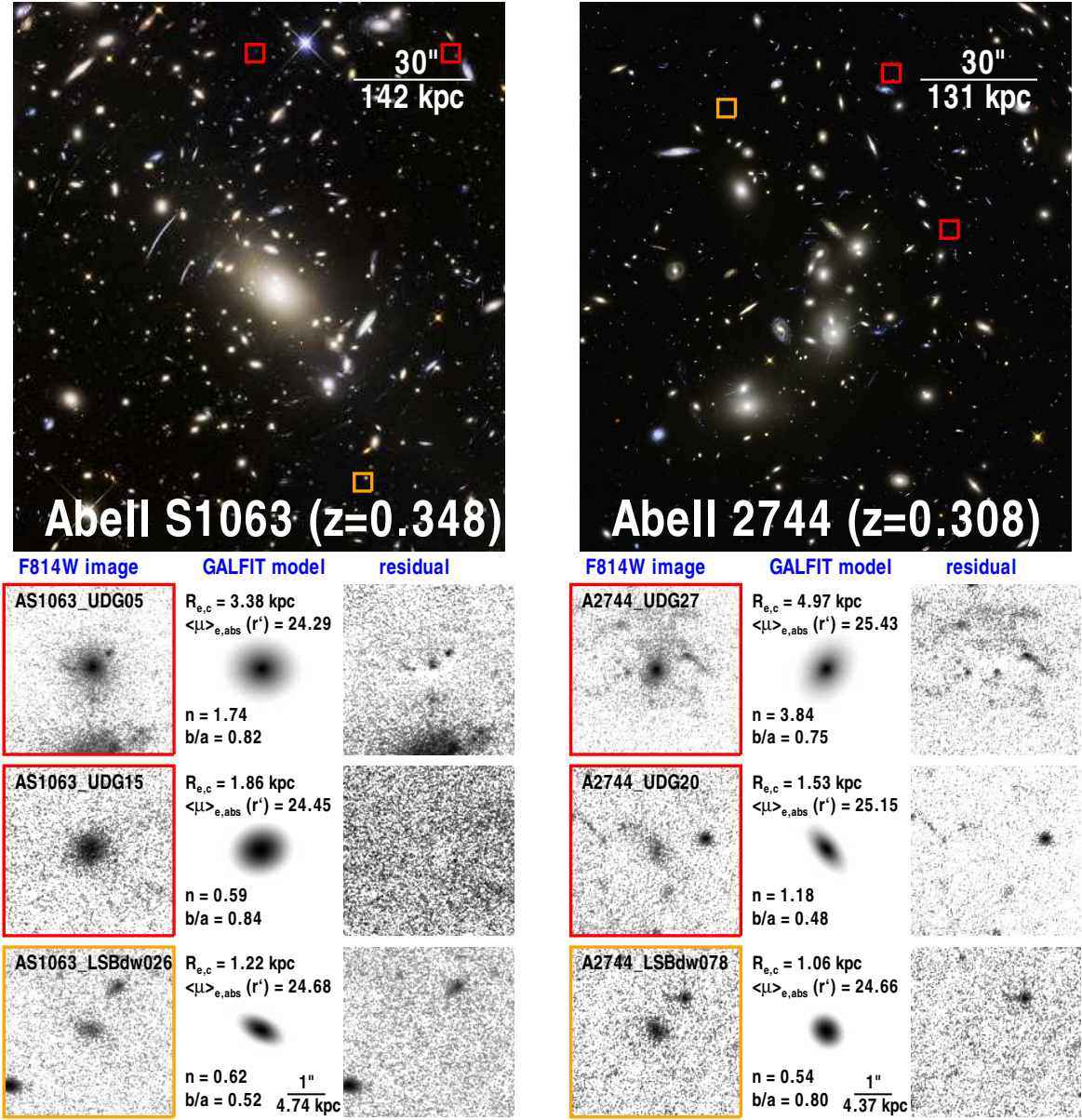} 
	\caption{(Upper panels) 
		Color images of 
		Abell S1063 (left) and Abell 2744 (right) HST fields.
		Red and yellow squares represent the position
		of example UDGs and LSB dwarfs, 
		respectively.
		(Lower panels) $4.5\arcsec\times4.5\arcsec$ zoom-in images of two UDGs and one LSB dwarf in each cluster.
		Left, middle, and right sections show
		F814W images, GALFIT models, and residual images after subtracting GALFIT models,
		respectively.
	}
	\label{fig_finder}
\end{figure*}

We used ACS/F814W($I$) and WFC3/F105W($Y$) images for Abell S1063 and Abell 2744 in the HFF \citep{lot16}.
 The effective wavelengths of the F814W and F105W filters for the redshifts 
 of Abell S1063 and Abell 2744 
 (6220 \AA~  and 8030 \AA)  correspond approximately to SDSS $r'$ and Cousins $I$ (or SDSS $i'$) in the rest-frame,  respectively. 
 This combination of filters is efficient for  the search of old stellar systems like UDGs and globular clusters in these galaxy clusters \citep{lee16}.
The HFF provides data for the central field of each cluster and the parallel fields at $\sim 6'.1$ east of Abell S1063 and at $\sim 5'.9$ west of Abell 2744.  Thus the parallel fields are close to, but still within the virial radius of each cluster.
We prepared the deep images drizzled with a pixel scale of $0\farcs03$.
The total exposure times are:
$T_{\rm exp} ({\rm F814W}) =$ 116,169s and $T_{\rm exp} ({\rm F105W}) =$ 67,341s  for Abell S1063, 
$T_{\rm exp} ({\rm F814W}) =$ 106,998s and $T_{\rm exp} ({\rm F105W}) =$ 66,141s  for the Abell S1063 parallel field,
$T_{\rm exp} ({\rm F814W}) =$ 104,270s and $T_{\rm exp} ({\rm F105W}) =$ 68,952s  for Abell 2744, and 
$T_{\rm exp} ({\rm F814W}) =$ 107,766s and $T_{\rm exp} ({\rm F105W}) =$ 67,329s  for the Abell 2744 parallel field. 
{\bf Figure \ref{fig_finder}} (Upper panels) displays color images of the HST fields for Abell S1063 and Abell 2744.

The full width at half-maximum (FWHM) values of the point sources in the images are  $\sim$3.0 pixels ($=0\farcs09$). 
These FWHM values  correspond to $\sim$430 pc for  Abell S1063 and $\sim$390 pc for Abell 2744.
Therefore the sources larger than these values can be detected as extended sources in the images
of the galaxy clusters.
As a reference for the
 background control, we used the  data for  the Hubble Extreme Deep Field (HXDF)
 (RA(2000)= 3$^h$ 32$^m$ 38$^s$.8, Dec(2000)= --27$^\circ$ $47'$ $28''$) 
provided by \citet{ill13}, as in the study of globular clusters, ultracompact dwarfs, and dwarf galaxies in 
Abell 2744 by \citet{lee16}.

\subsection{Selection of UDGs}

We searched for UDGs in the images of the target fields, considering mainly the effective radii and surface brightness, and secondarily the colors,  of the extended sources.
Our search procedure is similar to those in \citet{yag16, vander16}, which consists mainly of two steps: the first based on the application of SExtractor for source detection \citep{ber96},
and the second based on the application of GALFIT for parameter measurements \citep{pen10}. 

First, we ran 
SExtractor 
as a dual mode to detect extended sources and derive their photometry in the  images. 
We used only the sources detected in both F814W images and F105W images.
F814W images were used as a reference image for the dual mode photometry so that structural parameters of the detected sources are based on the F814W images.
We used SExtractor parameter values: DETECT\_MINAREA = 20 pixels, SEEING\_FWHM $= 0\farcs09$, DEBLEND\_NTHRESH = 32,  DEBLEND\_MINCONT = 0.01, BACKPHOTO\_TYPE = LOCAL and BACK\_SIZE = 32  pixels. 
We chose top-hat filters which are optimized for the detection of extended LSB objects, with  a low detection threshold, DETECT\_THRESH = $0.7\sigma$.
For photometry of the sources we used PHOT\_AUTOPARAMS (Kron factor, minimum radius) = (2.5, 3.5) for magnitudes, and (2.5, 1.75) for colors. 
We calibrated the instrumental magnitudes of these sources to  the AB system,
 following the STScI webpage\footnote{http://www.stsci.edu/hst/acs/analysis/zeropoints,
 http://www.stsci.edu/hst/wfc3/phot\_zp\_lbn}. 
Note that \citet{lee16} adopted Vega magnitudes for their photometry of the sources in Abell 2744.   AB magnitudes for F814W are 0.424 mag fainter  than Vega magnitudes: F814W(AB) = F814W(Vega)$+0.424$. We are using AB magnitudes in this study for comparison with other studies of the UDGs.

Slightly different parameters (e.g., major axis effective radii or circularized effective radii, surface brightness at the center or at effective radius) and values  were used for the selection of UDGs in the literature, as summarized in \citet{yag16} (see their Table 5).
We selected initial UDG candidates from the detected sources, using the generous selection criteria as follows:
stellarity (CLASS\_STAR) $<0.4$,
circularized effective radius $R_{\rm e,c, SE}$(FLUX\_RADIUS) $> 1$ kpc,
central surface brightness $\mu_{0,{\rm F814W, SE}}$(MU\_MAX) $>22.5$ \SBunit,
elongation parameter ($q=b/a$ where $a$ and $b$ are major and minor axes of the sources) $q>0.3$, $-0.5<$ F814W--F105W $<1.0$, and FLAGS $<4$.
Here we adopted a minimum value $R_{\rm e,c, SE}$ = 1 kpc, which is smaller than the UDG limits used in the literature.
The effective radii and central surface brightness values of the sources provided by SExtractor show good correlations with the values provided by GALFIT, but they show some 
differences, especially for multiple sources. 
Thus 
SExtractor values are used only for the initial selection of the UDG candidate.  
The numbers of these candidates are 521 and 304 
for Abell S1063 and its parallel field, and
352 and 295 
for Abell 2744 and its parallel field, respectively.

Second, we ran GALFIT to the images of the sources in the initial list of UDG candidates, deriving
effective radii and surface brightness at the effective radii of the sources. 
Before fitting of the UDG candidate images, we masked out neighboring sources around the target UDGs using the list of the sources detected with SExtractor and the segmentation maps. 
We derived point spread functions (PSFs) of the point sources in the images using PSFEx \citep{ber11}, which are used as an input for GALFIT measurements.
We performed the fitting of the surface brightness profiles with S\'{e}rsic index $n$ free. 
 Circularized effective radii ($R_{\rm e,c}$) were calculated from the major axis effective radii ($R_{\rm e}$) and elongation parameter, 
with $R_{e,c} = R_e \sqrt{q}$. 
We used these GALFIT values as the final 
values for the UDGs.

Finally we inspected the images of these galaxies, removing sources with artifacts,
multiple sources, sources close to the frame edges, gravitational lens arcs, and bright galaxies. 
Most of the finally selected UDGs show smooth structures, but some of them show substructures such as nuclei and disk features.
%

For the comparison of the UDGs at different redshifts, we transform the measured surface brightness at $z$ to the value at the rest frame ($z=0$).
The mean effective surface brightness 
 for the redshift $z$ ($ \langle\mu\rangle_{\rm e,z}(\lambda)$)  is given
in terms of the evolutionary correction $E(z)$ and the $K$-correction $K(z)$, so that the absolute mean effective surface brightness $ \langle\mu\rangle_{\rm e,abs}(\lambda)$ can be derived as follows ((Eq) 13 in \citet{gra05}):

\begin{equation}
\langle\mu\rangle_{\rm e,abs}(\lambda) = \langle\mu\rangle_{\rm e,z}(\lambda) - 10 {\rm log} (1+z) - E(z) - K(z).
\end{equation}

\noindent We derived $E(z)$ and $K(z)$ of the sources using GALAXEV for simple stellar populations with ages of 12 Gyrs \citep{bru03}.
The adopted values of the parameters for GALAXEV are as follows: the Chabrier stellar initial mass function \citep{cha03}, [Fe/H] $= -0.6$, and the age of 8.3 Gyrs for Abell S1063 and 8.7 Gyrs for Abell 2744.
The values of $E(z)$ and $K(z)$ in the F814W band are
--0.36 and +0.11 for Abell S1063, and 
--0.32 and +0.09 for Abell 2744, respectively.
If we adopt the age of 10 Gyrs for $z=0$,
the ages of Abell S1063 and Abell 2744 will be 6.3 Gyrs and 6.7 Gyrs, respectively,
and the values of $K(z)$ will be
+0.06 for Abell S1063, and +0.05 for Abell 2744.
We converted measured F814W magnitudes to SDSS $r'$-band using the values given by GALAXEV.



\begin{figure*}
	\centering
	\includegraphics[scale=0.7]{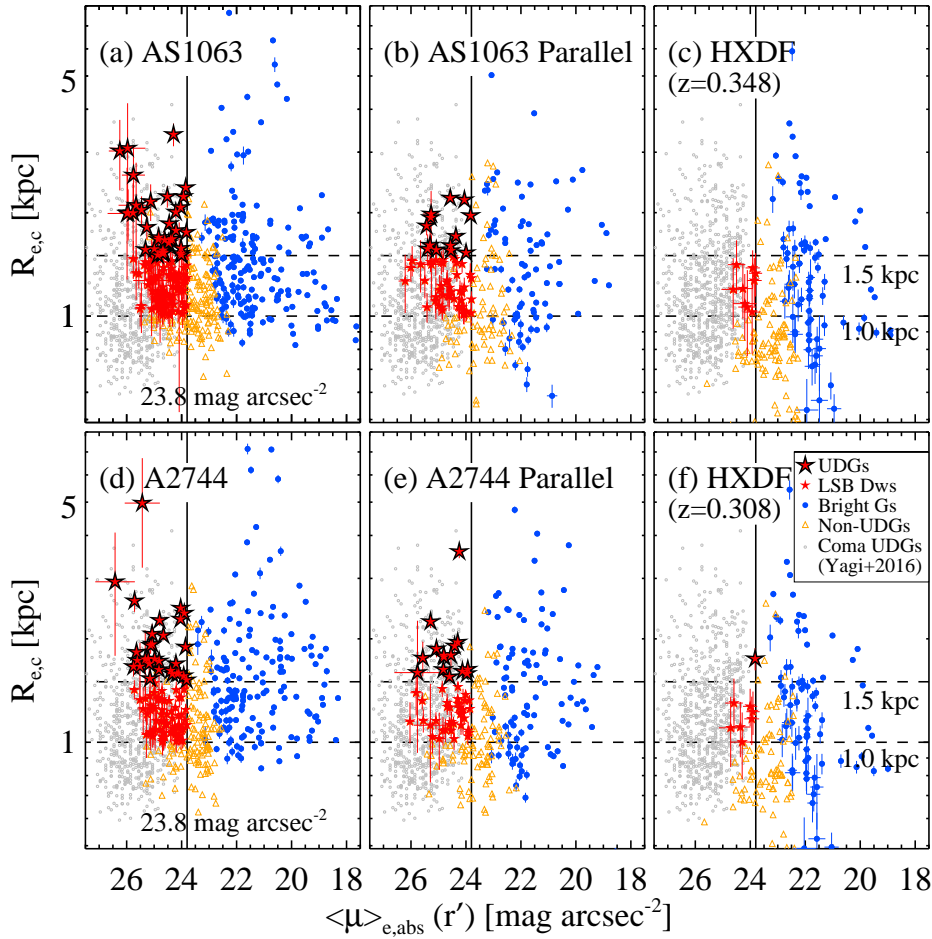} 
	\caption{
		Scaling relations of the galaxies in Abell S1063, Abell 2744, the parallel fields, and the HXDF, derived with GALFIT.
		Circularized effective radii ($R_{\rm e,c}$) versus absolute mean effective surface brightness (\absmue) 
		transformed to SDSS $r'$-band from F814W band: large red starlets for UDGs, small red starlets for LSB dwarfs, 
		blue circles for bright galaxies with high central surface brightness ($\mu_{\rm 0,F814W,SE} <22.5$ \SBunit), and yellow triangles 
		for non-UDGs (which were in the initial list of UDGs but were excluded later).
		The parameters for the HXDF galaxies are calculated for the redshifts of Abell S1063 and Abell 2744, respectively.
		Gray dots represent the UDGs in Coma \citep{yag16}. 
	}
	\label{fig_sel}
\end{figure*}

{\bf Figure \ref{fig_sel}} displays 
$R_{\rm e,c}$ 
versus 
\absmue of the galaxies
in Abell S1063, Abell 2744, the parallel fields, and the HXDF.
The parameters for the galaxies in the HXDF were derived for the redshifts of Abell S1063 and Abell 2744, respectively. Most of the galaxies detected in the HXDF are much more distant than the two clusters. Thus the parameters for the HXDF are useful only as a reference.
We also plotted the data for the Coma UDGs \citep{yag16} for comparison.

\begin{figure*}
	\centering
	\includegraphics[scale=0.7]{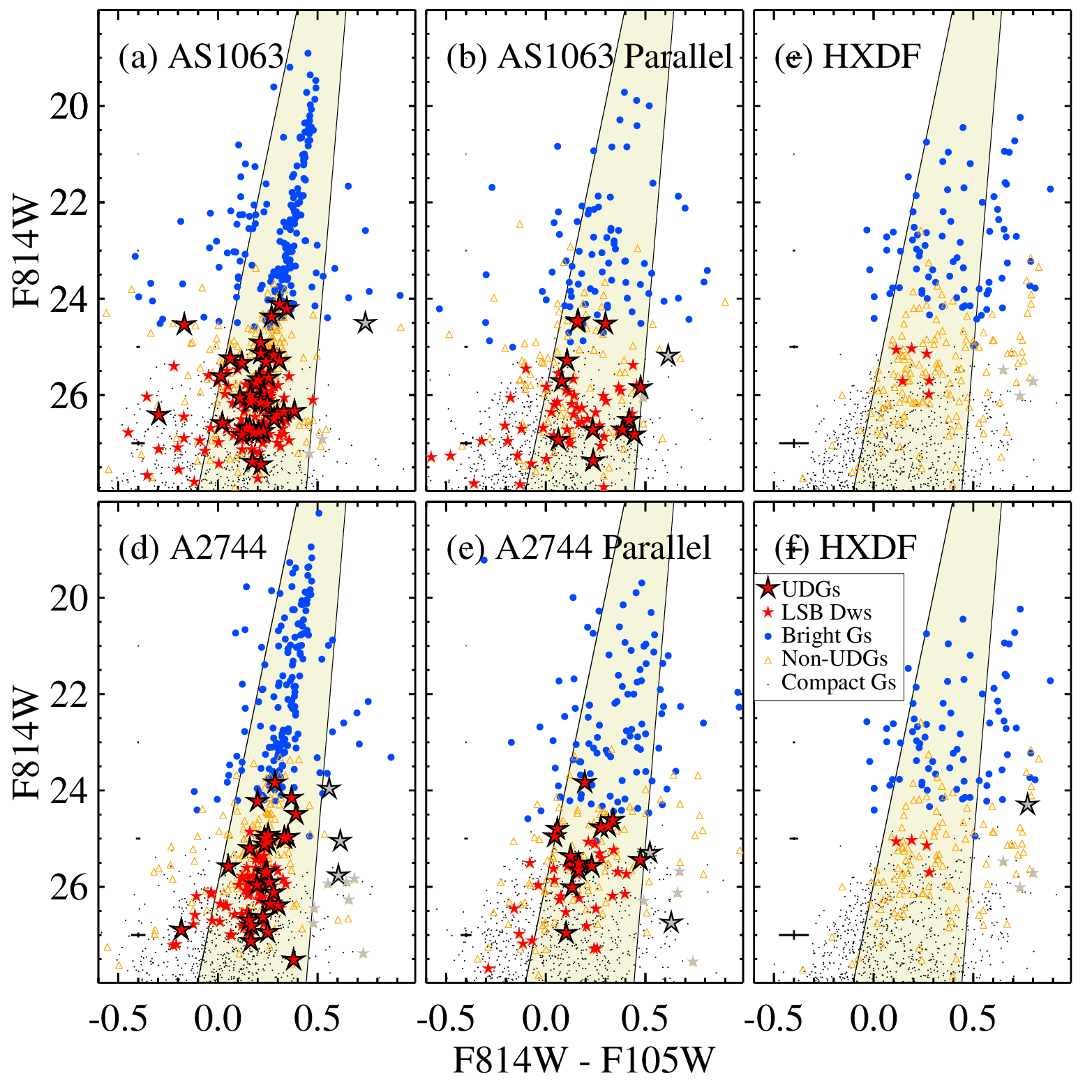} 
	\caption{CMDs of the UDGs and other galaxies in Abell S1063, Abell 2744, the parallel fields, and the HXDF.
		Symbols are same as in {\bf Figure \ref{fig_sel}},
		except for black dots denoting the compact galaxies with $R_{\rm e,c,SE} <1$ kpc here. 
		Errorbars in the left side represent the mean errors of magnitudes and colors.
		Solid lines represent the boundary for selecting the red sequence galaxies.
		Large and small gray starlets are, respectively, UDG and LSB candidates that are redder than the red sequence 
		and are removed from the UDG sample.
	}
	\label{fig_cmd}
\end{figure*}

\begin{figure*}
	\centering
	\includegraphics[scale=0.7]{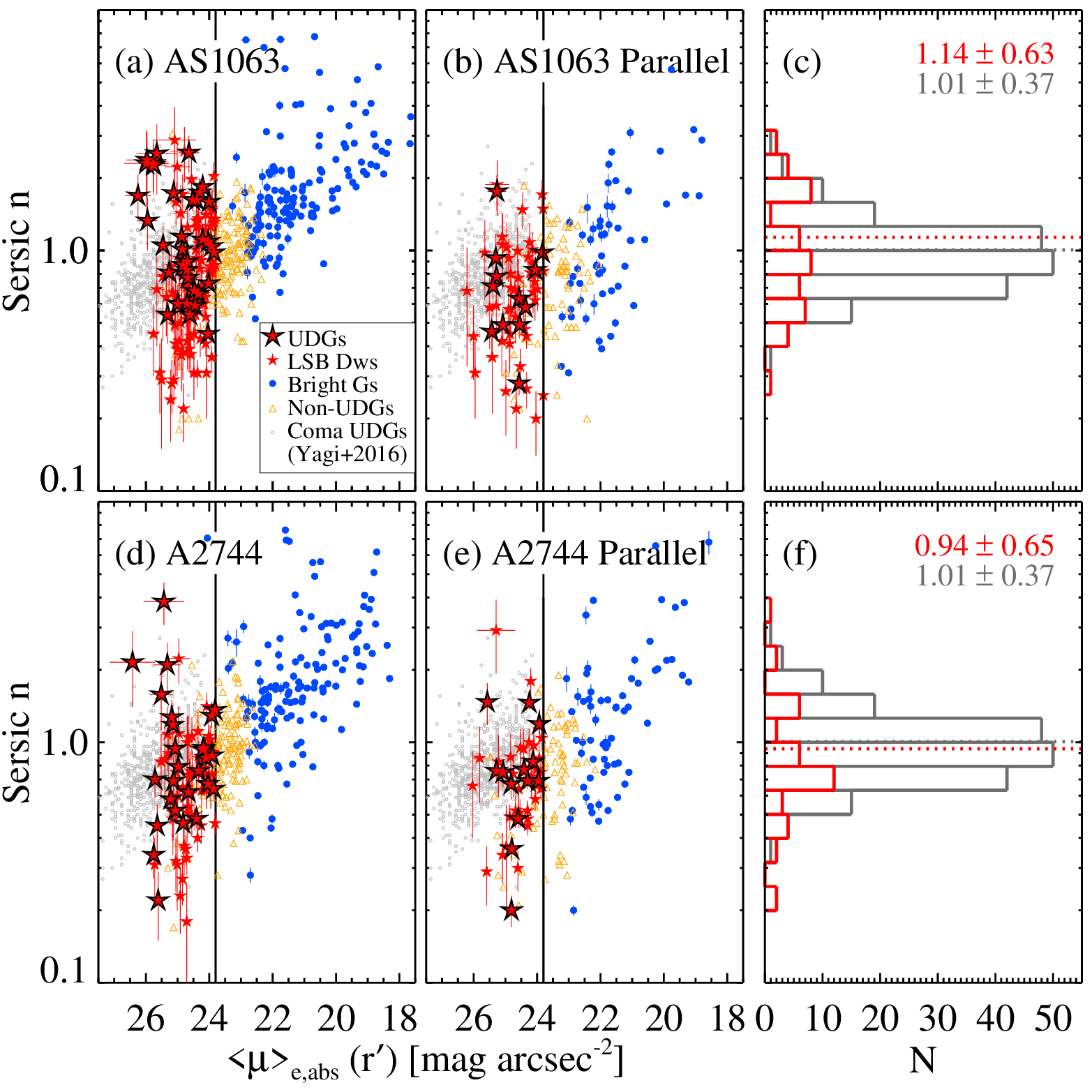} 
	\caption{
		(Left and middle panels) S\'{e}rsic index $n$ versus \absmue  
		for the UDGs and other galaxies in Abell S1063, Abell 2744, and the parallel fields. 
		Blue circles denote only the bright galaxies in the red sequence.
		Gray dots denote the Coma UDGs \citep{yag16}. 
		(Right panels) 
		Red line histograms are for the UDGs in Abell S1063 and Abell 2744, and gray line histograms are for the MW-sized UDGs in Coma. 
		The numbers and dotted lines represent mean values of S\'{e}rsic index $n$ for UDGs.
	}
	\label{fig_sersicn}
\end{figure*}

\begin{figure*}
	\centering
	\includegraphics[scale=0.7]{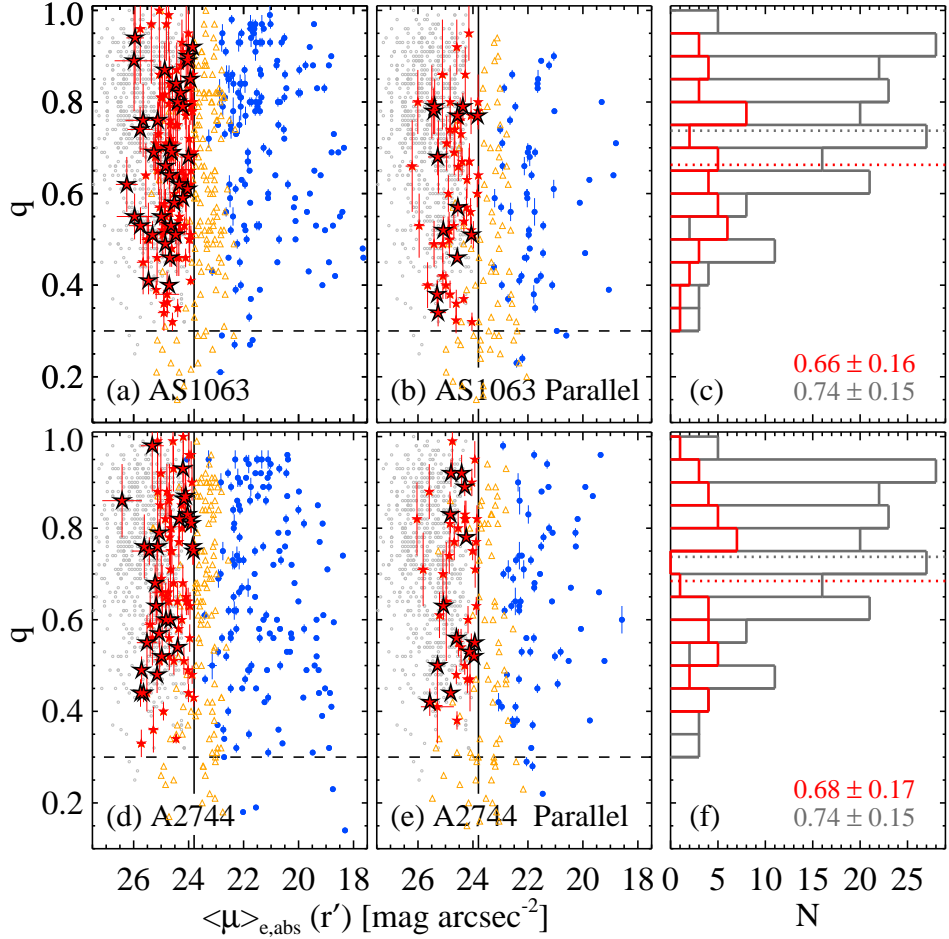} 
	\caption{
		Elongation parameter ($q=b/a$) versus \absmue (left and middle panels) and histograms of elongation parameter (right panels)
		for the UDGs and other galaxies in Abell S1063, Abell 2744, and the parallel fields. 
		Symbols are same as in {\bf Figure \ref{fig_sersicn}}.
	}
	\label{fig_elong}
\end{figure*}

\begin{figure*}
	\centering
	\includegraphics[scale=0.7]{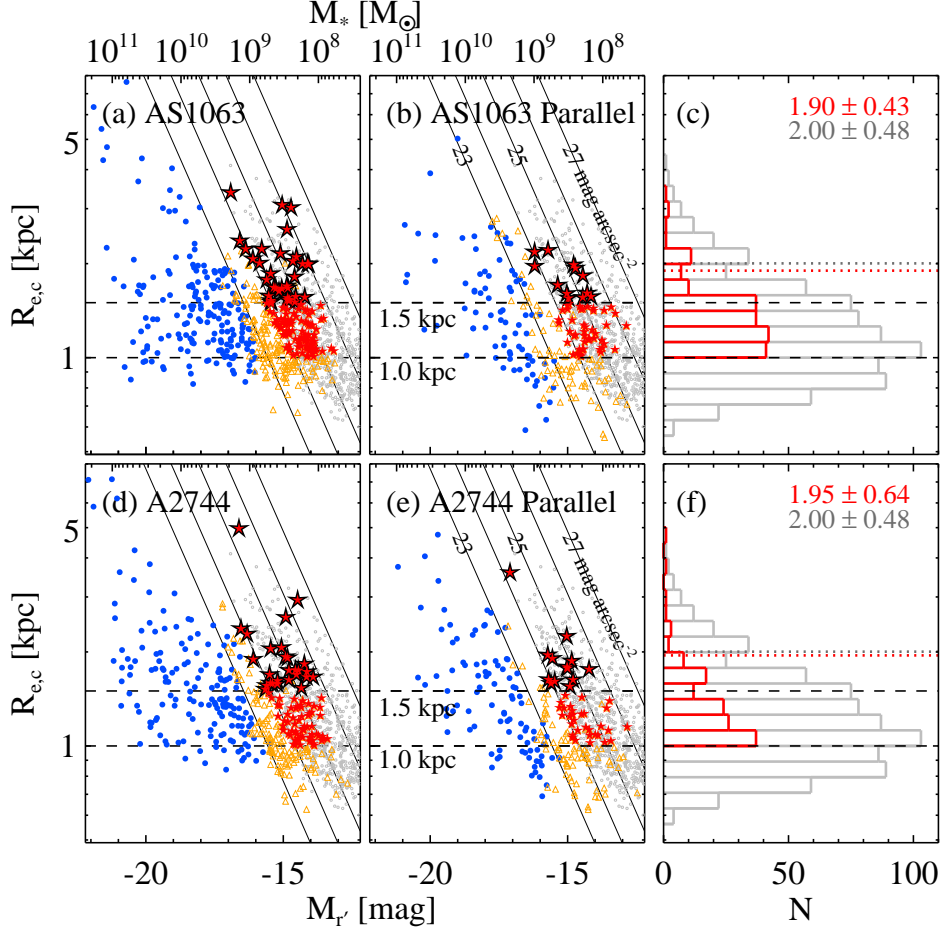} 
	\caption{
		(Left and middle panels) Circularized effective radii versus $r'$-band absolute magnitudes (lower X-axes) and stellar masses (upper X-axes) 
		of the UDGs and other galaxies in Abell S1063, Abell 2744, and the parallel fields. 
		Symbols are same as in {\bf Figure \ref{fig_elong}}. Solid lines represent iso-surface brightness magnitudes 
		of 
		\absmue = 23, 24, 25, 26, and 27 \SBunit 
		from left to right. 
		(Right panels) Red line histograms are for the UDGs and LSB dwarfs in Abell S1063 and Abell 2744, and gray line histograms for the UDGs in Coma.
		The numbers and dotted lines represent mean values of S\'{e}rsic index $n$ for UDGs.
	}
	\label{fig_sizemag}
\end{figure*}

\begin{figure*}
	\centering
	\includegraphics[scale=0.7]{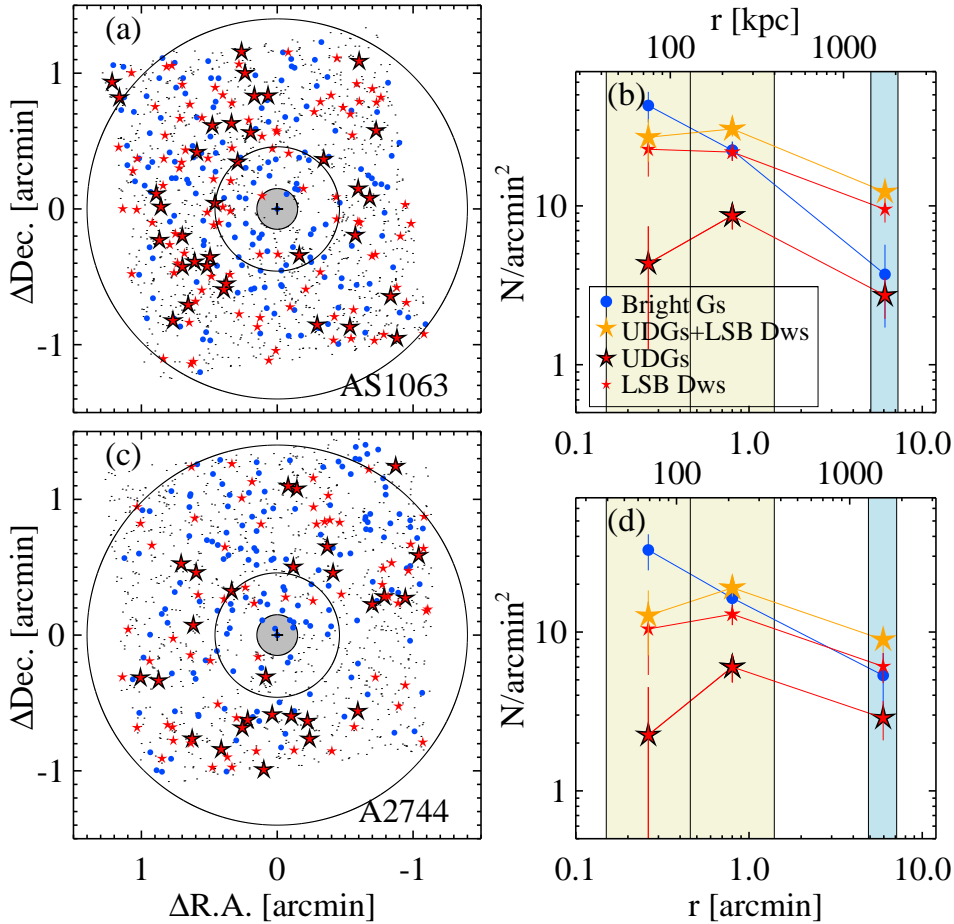} 
	\caption{Spatial distributions (left panels) and radial number density profiles corrected for detection completeness (right panels) of the UDGs (large red starlets), LSB dwarfs (small red starlets), and bright red sequence galaxies (blue circles) in Abell S1063 and Abell 2744. 
		Black dots denote the compact galaxies with $R_{\rm e,c,SE} <1$ kpc.
		The circles in the left panels represent the boundaries of the radial bins used for deriving radial number density profiles. The central gray circle denotes the region not used for the UDG survey.
		Yellow starlets in the right panels are for the sum of the UDGs and LSB dwarfs.
		Yellow and cyan regions represent the cluster and parallel fields, respectively.
	}
	\label{fig_map}
\end{figure*}

As the final UDGs (plotted by large red starlets), we selected the galaxies with $R_{\rm e,c} >1.5$ kpc, \absmue $>23.8$ \SBunit, and $q>0.3$,  
following the selection criteria adopted for the nearby galaxy clusters in \citet{vander16}. 
\citet{vander16} adopted a criterion, 
$\langle\mu\rangle_{\rm e, z=0.055} (r') >24.0$ \SBunit
for the nearby galaxies at the mean redshift of $z=0.055$, which corresponds to \absmue $>23.8$ \SBunit for $z=0$.
These galaxies are often called Milky Way (MW)-sized UDGs (Note that  
a slightly smaller limit, $R_{\rm e,c} >1.25$ kpc (or $R_{\rm e}>1.5$ kpc) is sometimes used for selecting the MW-sized UDGs in the literature \citep{van15,rom16b}).
Most of the UDGs in Abell S1063 and Abell 2744 have $R_{\rm e,c}=1.5-3$ kpc and \absmue $=23.8-26.5$ \SBunit.
In the figure, 
blue circles represent the bright normal galaxies with high central surface brightness $\mu_{0,{\rm F814W, SE}}<22.5$ \SBunit.

In addition, we selected the LSB dwarf galaxies with the same surface brightness range as, but smaller than, the UDGs: $1.0<R_{\rm e,c} <1.5$ kpc and 
 \absmue $>23.8$ \SBunit (plotted by small red starlets). Similar galaxies were included in the sample of Coma UDGs 
 in the studies of \citet{kod15,yag16}.
 \citet{yag16} called these galaxies as Subaru UDGs to distinguish from the MW-sized UDGs. 
The sample of LSB dwarf galaxies is used for the comparison with UDGs in our target clusters.
The locations of the new UDGs  and LSB dwarfs in Abell S1063 and Abell 2744 in {\bf Figure \ref{fig_sel}} are consistent with those of the Coma UDGs with higher surface brightness and larger sizes.

We marked, with yellow triangles in the figure, the galaxies that were included as the initial UDG candidates based on the SExtractor criteria, but were excluded later according to the GALFIT criteria.
They are mostly located between the domain of the  UDG plus LSB dwarfs and the domain of the bright galaxies. We call them as non-UDGs hereafter.

{\bf Figure \ref{fig_finder}} (Lower panels) displays zoom-in images of two UDGs and one LSB dwarf in each cluster. The left, middle, and right columns for each cluster
show F814W images, GALFIT model images, and the residual images after GALFIT model subtraction, respectively.
The numbers in the middle panels show the values of the
fitting parameters given by GALFIT.

\section{RESULTS}

\subsection{Color-Magnitude Diagrams of the UDGs}

In {\bf Figure \ref{fig_cmd}} we display the color-magnitude diagrams (CMDs) of the UDG candidates as well as other galaxies in Abell S1063, Abell 2744, the parallel fields, and the HXDF.

Following features are noted in this figure. First, both clusters show a strong red sequence, while the HXDF does not.
Second, the red sequences in the parallel field are much weaker than those in the cluster field, showing that the contributions due to background galaxies are larger in the parallel fields.
Third, no UDG candidates are found in the HXDF, while a small number of UDG candidates are seen in the parallel fields. This indicates that the UDG candicates in the parallel fields are cluster memebers.
In addition, a much more number of LSB dwarf candidates are seen in the parallel fields than in the HXDF, showing that most of them are cluster members.
Fourth, the UDGs and LSB dwarfs are located at the faint end of the red sequence, showing that they are mainly made  of old stars. A small number of the faint UDGs and LSB dwarfs in both clusters 
are bluer than the red sequence, indicating that they have some young stellar populations.
Fifth, the UDGs 
are fainter than the bright normal galaxies in the red sequence, and the F814W magnitude of the UDGs ranges from 24.0 to 28.0 mag. 

Thus most of the UDGs and LSB dwarfs are old stellar systems, while some show a recent activity of star formation.
 It is noted that a few UDG-like galaxies in the cluster and parallel fields 
 are redder than 
the red sequence (larger gray starlets).
Similarly a small number of LSB dwarfs are also redder than the red sequence (small gray starlets).  
They are probably background sources.
Therefore we excluded these galaxies in the final list of UDGs and LSB dwarfs. 


\subsection{The Census of the UDGs} 

Finally we select  
47 UDGs in Abell S1063 (35 UDGs in the cluster and 12 UDGs in the parallel field)  and
40 UDGs in Abell 2744 (27 UDGs in the cluster and 13 UDGs in the parallel field).
If we adopt the age of 10 Gyrs for $z=0$,
these numbers will be slightly changed:
53 UDGs in Abell S1063 (40 UDGs in the cluster and 13 UDGs in the parallel field)  and
42 UDGs in Abell 2744 (29 UDGs in the cluster and 13 UDGs in the parallel field).
Thus the numbers of UDGs in these two clusters are similar, which is consistent
with the expectation based on the similarity in the cluster mass.  We use these UDGs for the following analysis.

Also we select 96 and 47 LSB dwarf candidates in Abell S1063 and its parallel field (143 in total), and 
62 and 31 such sources in Abell 2744 and its parallel field (93 in total), 
respectively.
On the other hand, we see 
6 and 4 LSB dwarf-like galaxies  in the HXDF for the redshifts of Abell S1063 and Abell 2744, respectively.
The area ratios 
of the cluster and parallel fields with respect to that of the HXDF are 0.990 for Abell S1063, 0.989 for its parallel field, 
1.288 for Abell 2744, and 0.986 for its parallel field, respectively.
Thus the net numbers of the LSB dwarf galaxies in Abell S1063 and its parallel field are, respectively,  
90 and 41 (131 in total), 
subtracting the contribution of the background sources based on the HXDF.
Similarly we derive the net numbers of the LSB dwarf galaxies in Abell 2744 and its parallel field are, respectively,  
57 and 26 (83 in total). 

Tables 1 and 2 list the catalogs of UDGs and LSB dwarfs in Abell S1063,
and Tables 3 and 4 list the catalogs of UDGs and LSB dwarfs in Abell 2744, respectively.
In the tables we include the following information: IDs, RA(J2000), Dec.(J2000), major axis effective radii,  effective surface brightness, F814W magnitudes, (F814W--F105W) colors, S{\'e}rsic index $n$, elongation parameter, circularized effective radii, and absolute mean effective surface brightness of the galaxies.

\subsection{Structural Parameters of the UDGs}

We compare structural parameters of the UDGs in Abell S1063 and Abell 2744 with those in Coma \citep{yag16}. 
In {\bf Figure  \ref{fig_sersicn}} 
we plotted the S\'{e}rsic index $n$ versus \absmue of the UDGs (large red starlets) and LSB dwarfs (small red starlets) in Abell S1063 and Abell 2744 and their parallel fields and the Coma UDGs (gray dots). We also plotted the bright normal galaxies in the red sequence of Abell S1063 and Abell 2744 (blue circles) and the non-UDGs 
(yellow triangles).

The bright red sequence galaxies are mostly the members of each cluster and can be considered to be simple stellar populations of old age. 
In the right panels of the figure we show the histograms of S{\'e}rsic index $n$ for the UDGs in Abell S1063 and Abell 2744 in comparison with those of the Coma UDGs.
We selected only the MW-sized UDGs in the Coma UDG sample for comparison of the histograms.

It is seen that the values of $n$ for most UDGs in Abell S1063 and Abell 2744 are smaller than three, while some of the bright red sequence galaxies
have $n>3$.
The distributions of $n$ in these two clusters are similar to those of the MW-sized UDGs in Coma. The mean values,
 $\langle n \rangle=1.14\pm0.63$ for Abell S1063 and
$\langle n \rangle =0.94\pm0.65$ for Abell 2744,
are similar to that of the Coma UDGs,
$\langle n \rangle=1.01\pm0.37$. 
Thus the radial surface brightness profiles of the UDGs in Abell S1063 and Abell 2744 are mostly fit by an exponential profile, similar to the case of UDGs in Coma and other nearby galaxy clusters \citep{yag16,rom16b,vander16}.

{\bf Figure \ref{fig_elong}} 
displays the elongation parameter $q$  
versus \absmue (left and middle panels), of the same galaxies as before, 
and the histograms of elongation parameters for the same galaxies as in {\bf Figure \ref{fig_sersicn}}.
Most of the UDGs in Abell S1063 and Abell 2744 have the elongation parameters larger than 0.4.
The mean values of the elongation parameters for the UDGs in Abell S1063 and Abell 2744 are 
$\langle q \rangle=0.66\pm0.16$ and $0.68\pm0.17$,
which are similar to the value of the MW-sized Coma UDGs, $\langle q \rangle=0.74\pm0.15$.
Structural parameters ($n$ and $q$) of the LSB dwarfs in Abell S1063 and Abell 2744 show similar features to those of the UDGs.

\subsection{Stellar Masses of the UDGs}


In {\bf Figure \ref{fig_sizemag}} 
we plotted the effective radii versus $M_{r'}$ magnitudes of the galaxies in the same fields as before.
We derived a relation between the stellar mass and $M_{r'}$ magnitudes for simple stellar populations, for
the same GALAXEV parameters as described in Section 2.
We overlayed the values for
corresponding stellar masses in the upper axis of each panel. 
It shows that the absolute magnitudes of most UDGs in Abell S1063 and Abell 2744 range from $M_{r'} = -14.0$ to $-17.0$ mag,
and similarly their stellar mass varies from $M_* = 10^8$ to $10^9~ M_\sun$.  
The LSB dwarfs in Abell S1063 and Abell 2744 cover lower ranges than the UDGs: $M_{r'} = -13.0$ to $-15.5$ mag,
and $M_* = 5 \times 10^7$~ to~ $5\times 10^8~ M_\sun$.
The slanted solid lines represent the varying surface brightness, \absmue = 23, 24, 25, 26, and 27 \SBunit (from left to right).
For given surface brightness, the larger the UDGs are, the brighter (more massive) they are. The stellar masses of the UDGs are much lower than those of the bright red sequence galaxies.

In the right panels of {\bf Figure \ref{fig_sizemag}} 
we plotted the histograms of effective radii for Abell S1063 and Abell 2744 in comparison with that of the Coma UDGs \citep{yag16}. 
The mean values,
$\langle R_{\rm e,c}\rangle=1.90\pm0.43$ kpc for Abell S1063 and
$\langle R_{\rm e,c}\rangle=1.95\pm0.64$ kpc for Abell 2744,
are similar to that of the MW-sized UDGs in Coma,
$\langle R_{\rm e,c}\rangle=2.00\pm0.48$ kpc. 
The histograms of the combined sample of UDGs and LSB dwarfs are similar to those of the large UDGs in Coma.

\subsection{Spatial Distributions of the UDGs}

{\bf Figure \ref{fig_map}} (left panels) displays the spatial distribution of the UDGs, LSB dwarf galaxies, and bright red sequence galaxies in Abell S1063 and Abell 2744 fields.
The circles in the left panels represent the boundaries of the radial bins used for deriving radial number density profiles.
We excluded the central circular region in the UDG survey, because of high surface brightness of the central galaxy.

We derived the radial number density profiles of the galaxies, 
subtracting the background contribution using the data for the HXDF. 
As the center for the radial profiles we adopted the center of the cD galaxy in Abell S1063, and the center of CN-1 in Abell 2744. Note that \citet{lee16} adopted the center of CN-2 at the south-east of CN-1, which is the brightest galaxy in the HST field of Abell 2744,  for deriving radial number density profiles of the point sources (mostly globular clusters and ultra compact dwarfs). In this study of UDGs, we chose CN-1, which is closer to the center of the southern core in Abell 2744. 
We plotted the results in the right panels of the figure.

\begin{figure}
\centering
\includegraphics[scale=0.5]{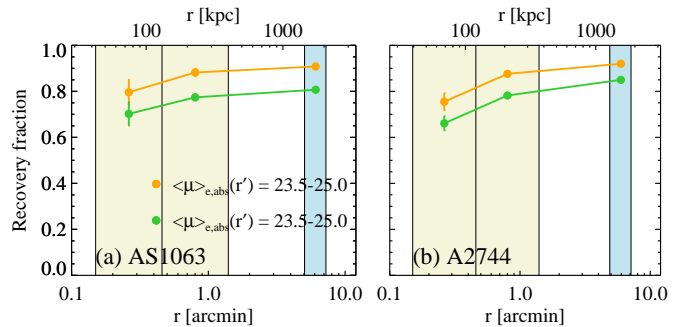} 
\caption{
Completeness of galaxy detection  with respect to projected clustercentric distance for Abell S1063 (a) and Abell 2744 (b). Yellow and green lines are for \absmue = 23.5 -- 25.0 \SBunit, and 
 \absmue = 25.0 -- 26.5 \SBunit, respectively.
}
\label{fig_recovery}
\end{figure}

We estimated the completeness of our UDG and LSB dwarf detection using a mock galaxy experiment.
We generated images of mock galaxies with structural parameters similar to those of the UDGs and LSB dwarfs onto the original images, using IRAF/ARTDATA. 
Then we repeated the same search procedure as used for UDG and LSB dwarf detection. Finally we derived a radial profile of completeness (recovery fraction) as a function of clustercentric distance for each cluster,
as shown in {\bf Figure \ref{fig_recovery}} for two magnitude ranges:
\absmue $=23.5-25.0$ and $25.0-26.5$ \SBunit.
We corrected the radial number density profiles of the sources for completeness using this result.

A few interesting features are distinguishable in {\bf Figure \ref{fig_map}}.
First, the radial number density profiles of the UDGs, LSB dwarfs, and bright red sequence galaxies in both clusters 
keep increasing as the clustercentric distance decreases in the outer region at 100 kpc $ <r<2$ Mpc. This shows that these galaxies are mostly indeed the members of each cluster. The slopes of the UDGs and LSB dwarfs are flatter than that of the bright galaxies in Abell S1063, while they are similar in Abell 2744. 
Second, 
the radial number density profiles of the UDGs and LSB dwarfs in both clusters show a drop or a flattening in the central region at $r<100$ kpc, while that of the bright galaxies keeps increasing in the central region. 
Thus the relative number density of the UDGs plus LSB dwarfs with respect to that of the bright galaxies is relatively lower in the central region than in the outer region. 
%
%
%
Third, spatial distributions of the UDGs are more inhomogeneous than those of the bright red sequence galaxies. This is a very interesting feature, which may be related with the origin of the UDGs. Inhomogeneous distributions of the UDGs indicate that UDGs are relatively new comers that came from outside the galaxy clusters so their distribution is not yet dynamically virialized. 
Further studies with a larger sample of galaxy clusters or with simulations of UDGs are needed to investigate this issue.


\section{DISCUSSION} 

\subsection{Total Masses of the UDGs}

The dynamical mass of the UDGs provides a critical information to understand the origin of the UDGs. However, it is difficult to determine the dynamical mass of the 
LSB galaxies like UDGs so
dynamical estimates of the UDG mass are available only for a few UDGs: Dragonfly 44 in Coma \citep{van16}, VCC 1287 in Virgo \citep{bea16},
and UGC 2162, the nearest UDG in the M77 group \citep{tru17}. 

\citet{zar08} suggested a new method to estimate the total mass of stellar systems without the kinematic measurements, which is based on the fundamental manifold.
A kinematic term of a stellar system 
($V= \sqrt{\sigma_v^2 + v_r^2/2}$~ where $\sigma_v$ is the line-of-sight velocity dispersion and $v_r$ is the rotational velocity) 
can be estimated, if the values of surface brightness and effective radii are known in the fundamental manifold. 

We estimate the total mass 
of the UDGs in Abell S1063 and Abell 2744, following this method that \citet{zar17} applied to the MW-sized UDGs in Coma \citep{van15} and the dwarf galaxies in Fornax clusters \citep{mun15}.
The galaxy scaling relations are given in terms of $V$, the mass-to-light ratio, $\Upsilon_e$, and the mean surface brightness, $I_e$,  within the effective radius, $R_{\rm e,c}$ \citep{zar08,zar17}:
\begin{equation}
\begin{split}
{\rm log} \Upsilon_e = 0.24 ({\rm log} V)^2 + 0.12 ({\rm log} I_e )^2 - 0.32 {\rm log} V \\ - 0.83 {\rm log} I_e - 0.02 {\rm log} V I_e + 1.49
\end{split}
\end{equation}
and 
\begin{equation}
{\rm log} R_{\rm e,c} = 2 {\rm log} V - {\rm log} I_e - {\rm log} \Upsilon_e - 0.75.
\end{equation}

\noindent Combining these two equations leads to
${\rm log} V = f(I_e , R_{\rm e,c} ).$
Thus we can derive the values of $V$ 
 and $\Upsilon_{\rm e}$ from $I_{\rm e}$ and $R_{\rm e,c}$. Here we assume $v_r= 0$ and $V=\sigma_v$ for velocity dispersion-supported systems. 

Then we estimate the enclosed mass within the 3D half-light radius $R_{1/2}$, 
using the formula for velocity dispersion-supported galaxies given by \citet{wol10}, 

\begin{equation}
M(<R_{1/2} ) = 4 \sigma_v^2 R_{\rm e,c} /G = 930 (\sigma_v/ {\rm km~ s^{-1}} )^2 (R_{\rm e,c}/{\rm pc}),
\end{equation}

\noindent where  $R_{1/2}$ is related with the 2D effective radius by $R_{1/2} = 4 R_{\rm e,c}/3$.
Finally the total mass (virial mass) of the galaxies can be estimated from a comparison of the enclosed mass with the NFW mass profiles for given concentration parameters \citep{nav97,lud16}. 

\begin{figure}
\centering
\includegraphics[scale=0.95]{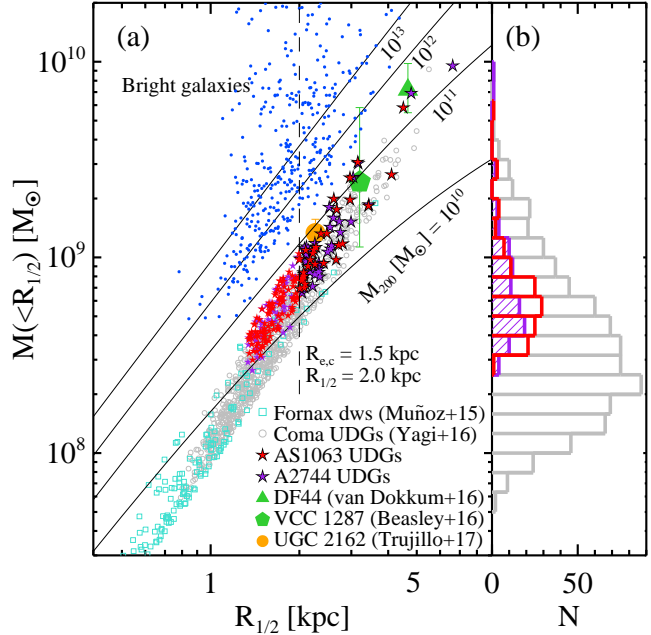} 
\caption{(a) 
Enclosed mass $M(<R_{1/2})$ versus 3D effective radii $R_{1/2}$ of the UDGs 
(large starlets) and LSB dwarfs (small starlets) in  Abell S1063 (red) and Abell 2744 (violet) in comparison with Fornax dwarfs (turquoise squares) 
 and Coma UDGs (gray circles). 
  Dragonfly 44  and VCC 1287 
  are  marked by a green triangle and a green pentagon, respectively 
(based on the observed values in the references \citep{van16,bea16}). UGC 2162, the nearest UDG, \citep{tru17} is also plotted by
a yellow circle.
Note that $R_{1/2}$ denotes 3D half-light radii, given by $R_{1/2} = 4 R_{\rm e,c}/3$.
Solid lines represent the NFW profiles for the total mass $M_{200} = 10^{10}, 10^{11}, 10^{12}$ and $10^{13} M_\sun$. 
Blue circles denote the bright red sequence galaxies in Abell S1063 and Abell 2744.
(b) Distributions of the enclosed mass of the UDGs and LSB dwarfs in Abell S1063 (red), Abell 2744 (violet), and Coma (gray).
}
\label{fig_mass}
\end{figure}

{\bf Figure \ref{fig_mass}} displays the enclosed mass versus the 3D half-light radii ($R_{1/2}$) of the UDGs, LSB dwarfs, and bright red sequence galaxies in Abell S1063 and Abell 2744. 
For comparison we also plotted the results for the UDGs in Coma \citep{yag16}
 and 
 the dwarf galaxies with
 \meanmue$ >23.8$ \SBunit in Fornax selected from the catalog given by \citet{mun15}. 
  Most of the Fornax dwarf galaxies in \citet{mun15} have effective radii smaller than 1 kpc and only a small number of them have effective radii larger than 1.5 kpc.  
We also plotted the data for
Dragonfly 44, VCC 1287, and UGC 2162. \citet{tru17} provided an enclosed mass within 5 kpc for UGC 2162. 
We multiplied it by a factor of 0.3 (the ratio of the radii $=1.7/5$) to estimate the enclosed mass for $R_{\rm e,c}=1.7$ kpc. Note that the relations based on the NFW density profiles for $M_{200}=10^{10}$, $10^{11}$, $10^{12}$, and $10^{13} M_\sun$ (with concentration parameter $c=$12.5, 10.6, 8.7, and 6.9 in \citet{lud16}) are displayed by solid curved lines.
The distributions of the enclosed mass of UDGs and LSB dwarfs in Abell S1063, Abell 2744, and Coma are shown in the right panel of the figure.

A few distinguishable features are noted in this figure.
First, the distributions of the enclosed mass of the UDGs in Abell S1063 and Abell 2744 are similar, and they are overlapped with the high mass part of the Coma UDGs. 
The enclosed masses of most UDGs range from $M(<R_{1/2} ) = 6 \times 10^8 M_\sun$~ to $3 \times 10^9 M_\sun$, and three largest UDGs have much higher masses, $6 \times 10^9 M_\sun$~ to $10^{10} M_\sun$.
Second, the larger the UDGs are, the higher enclosed mass they have.
Third, most of the UDGs in Abell S1063 and Abell 2744 have total masses
$M_{200} = 10^{10}- 10^{11} M_\sun$.
The LSB dwarfs have slightly lower masses,
$M_{200} = 1- 8\times 10^{10} M_\sun$. 
Fourth, only nine of the UDGs in these clusters have total masses larger than $M_{200} = 10^{11} M_\sun$. Six of them have total masses and sizes similar to VCC 1287, and three of them (AS1063\_UDG05, A2744\_UDG27, and  A2744\_UDG04 (located in the parallel field))
have total masses and sizes similar to those of Dragonfly 44 that is an proto-example of massive UDGs in Coma. 



\subsection{Total Numbers of UDGs and the Masses of their Host Systems}

From the study of UDGs in nearby galaxy clusters, \citet{vander16} found that the total number of UDGs in nearby galaxy clusters shows a strong correlation with the virial mass ($M_{200}$) of their host clusters. They fit the data with a power law, N(UDG) $= M_{200}^\alpha$, obtaining $\alpha=0.93\pm 0.16$.
This correlation is also seen in the expanded sample including other galaxy clusters and groups \citep{rom16a,jan16}.

From the numbers of the detected UDGs, we estimate the total numbers of UDGs in Abell S1063 and Abell 2744.
We counted the number of UDGs 
within the virial radius of each galaxy cluster.
Considering the radial number density profiles corrected for completeness
for each galaxy cluster, we derive
N(UDG)$=770\pm114$ 
for Abell S1063, and  
N(UDG)$=814\pm122$ 
for Abell 2744.

\begin{figure}
	\centering
	\includegraphics[scale=0.95]{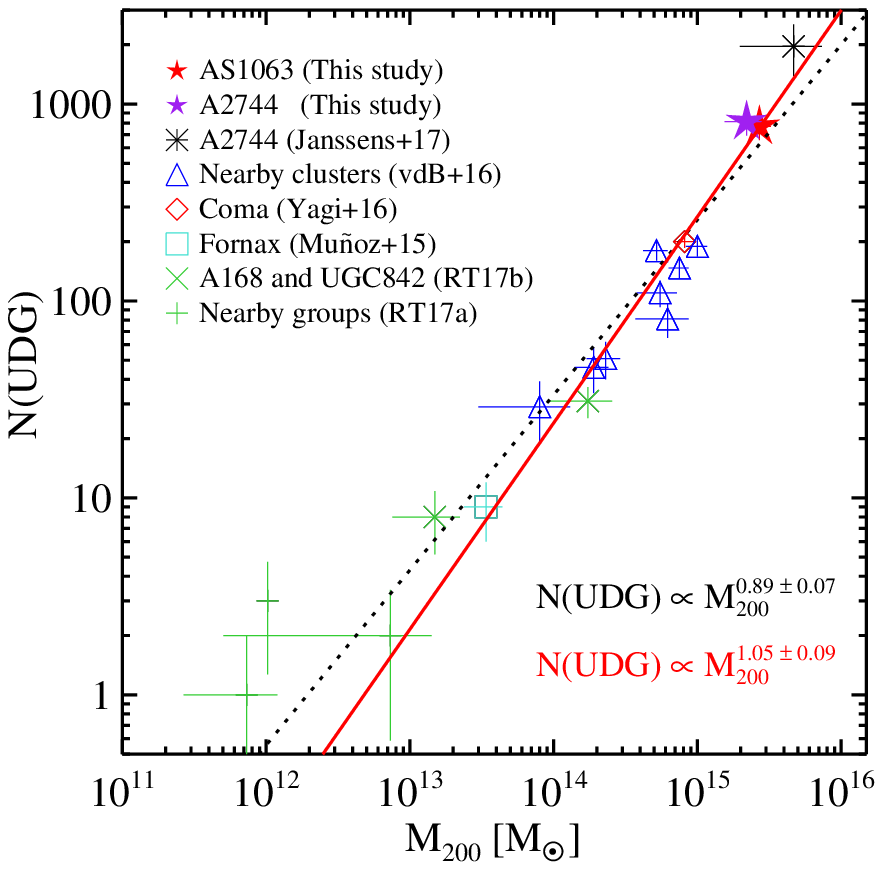} 
	\caption{ Total numbers of UDGs vs total mass of their parent galaxy clusters for Abell S1063 (red starlet), Abell 2744 (violet starlet) and other clusters in the literature \citep{mun15,yag16,vander16,jan16, rom16a,rom16b}.
		Dashed line and solid line represent the power law fitting for all data and for massive systems with $M_{200}>10^{13} M_\sun$, respectively. For fitting we used the values in this study in the case of Abell 2744.
	}
	\label{fig_nmass}
\end{figure}

In {\bf Figure \ref{fig_nmass}}
we plotted these results  in comparison with the previous results in the literature compiled by \citet{rom16a,jan16}: nearby galaxy clusters  \citep{vander16}, Coma \citep{yag16}, Fornax \citep{mun15}, Abell 168 and UGC 842 \citep{rom16b}, three Hickson compact groups \citep{rom16a}, and Abell 2744 \citep{jan16}.
Note that we adopted the mass of Abell 2744, $M_{200}=2.2^{+0.7}_{-0.6} \times 10^{15} M_\sun$, as described in Section 1, which is smaller than the value $M_{200}=5 \times 10^{15} M_\sun$ used in \citet{jan16}.
Abell S1063 and Abell 2744 are the most massive clusters in this sample so that they are precious targets to extend the range of virial mass in the study of this relation.

In the figure, the data for the UDGs in Abell S1063 and Abell 2744 in this study appear to be located at the upper end of the previous data.  Fitting the data for all UDGs with a power law
(we use the values in this study in the case of Abell 2744), we obtain  $\alpha=0.89\pm0.07$. 
 This value for the power law index is very similar to that given by \citet{rom16a}, $\alpha=0.85\pm0.05$, and is consistent with the value presented by \citet{vander16}, $\alpha=0.93\pm0.16$. 
Thus the two clusters in this study follow well the power law given by lower mass systems.
It is interesting that the data for the galaxy group mass of $M_{200}=10^{12}$ to the massive cluster mass of $3 \times 10^{15} M_\sun$ are represented remarkably well by the power law.

The derived value of the power law index, $0.89\pm0.07$, is slightly smaller than one. 
It indicates that the formation (or survival) efficiency of UDGs may be higher in the lower mass systems \citep{rom16a}.
 Based on the lower power law index value,
$\alpha=0.85\pm0.05$ derived in their study,  
\citet{rom16a} suggested a scenario that UDGs are dwarf galaxies and that the progenitors of today's UDGs are formed in the low density field, are processed in galaxy groups, and then some of them are disrupted during the infall to galaxy clusters. 

However, the data for the low mass end have much larger errors than those for the high mass systems, because the numbers of UDGs in the Hickson compact groups presented by \citet{rom16a} are small. 
If we fit the data for massive systems with $M_{200}>10^{13} M_\sun$ in the sample, we derive  $\alpha = 1.05\pm0.09$.  
 This value is close to one, implying that the formation (or survival) efficiency of UDGs depends little  on the their host mass for $M_{200}>10^{13} M_\sun$.
 It is needed to study more UDGs in the galaxy groups with $M_{200}<10^{13} M_\sun$ to investigate the flattening of the slope in the low mass range.

It is noted that the data for Abell 2744 given by \citet{jan16} shows a slightly larger deviation from the power law fit than our result, if the same value for the cluste mass is adopted. If it is real, Abell 2744 has an excess of UDGs compared with lower mass systems.
However, the total number of UDGs in Abell 2744 derived in this study, $N=814\pm122$, 
is about a factor of two 
smaller than the value given by \citet{jan16},  $N=1961\pm577$. 
The cause for this difference is not clear.
The numbers of the UDGs we detected in Abell 2744 and its parallel field
are  27 and 13, 
 respectively (and none in the HXDF).
On the other hand, \citet{jan16} found 41 and 35 UDGs in Abell 2744 and its parallel field, and  10 UDGs in the HXDF.
Their sample includes more UDGs larger than $R_{\rm e,c} >3$ kpc, and some UDGs with lower surface brightness, reaching \absmue = 26.3 \SBunit. We had also similar sources in our initial list of UDG candiates, but most of them were removed in the visual inspection step.  
Therefore about a half of the difference between the two studies may be
due to the difference in the detected numbers, and another half may be due to the difference in the total number estimation procedure.



\subsection{Origin of UDGs}

Here we discuss the primary results of the UDGs in Abell S1063 and Abell 2744 derived in the previous sections, in relation to the scenarios for the origin of the UDGs: an inflated dwarf galaxy scenario \citep{yoz15,amo16,bea16,rom16b,dic17} 
and a failed $L_{*}$ galaxy scenario\citep{van15,kod15,van16}.

{\it Radial Number Density Profiles of the UDGs}: 
 We found two interesting results on the radial distributions of the UDGs in Abell S1063 and Abell 2744: 
a) a similarity between  the radial number density profiles of the UDGs, the LSB dwarfs,
and the bright red sequence galaxies in the outer region at 100 kpc $<r<2$ Mpc of the galaxy clusters, and 
b) the number density ratios of the UDGs plus LSB dwarfs with respect to the bright red sequence galaxies that is relatively lower in the central region (at $r<100$ kpc) than in the outer region (at $r>100$ kpc) of the galaxy clusters. 
These results imply the following points.
First, the UDGs and the LSB dwarfs are less massive than the bright red sequence galaxies so that they cannot survive 
as long as the bright galaxies in the central region of the cluster.
Second, the UDGs and the LSB dwarfs are vulnerable to the harsh environments in the galaxy clusters so a significant fraction of them are disrupted in the central region of the clusters.   

{\it Total Masses of the UDGs:} 
A small number of the UDGs in Abell S1063 and Abell 2744 have high total masses exceeding $M_{200} = 10^{11} M_\sun$.
Six of them have total masses and sizes similar to those of VCC 1287, but three of them have
total masses and sizes similar to those of Dragonfly 44 in Coma.
Dragonfly 44  (with $ M=8 \times 10^{11} M_\sun$ \citet{van16}) is known to be one of the most massive UDGs in the local universe, being used as an example for UDGs with the $L_{*}$ galaxy origin. 
Thus the most massive UDGs in these galaxy clusters can be failed $L_*$ galaxies.
On the other hand, a majority of the UDGs in Abell S1063 and Abell 2744 have total masses smaller than $M_{200} = 10^{11} M_\sun$. However, they are more massive than $M_{200} = 10^{10} M_\sun$.
Thus they correspond to the upper end in the mass distribution of Coma UDGs (see {\bf Figure \ref{fig_mass}}). This result is consistent with
the inference based on the radial number density distribution in the previous section.

From these results, we conclude that a majority of the UDGs in Abell S1063 and Abell 2744 are relatively massive dwarf galaxies, supporting the dwarf galaxy origin hypothesis. A small number of the UDGs in these galaxy clusters have masses and sizes similar to Dragonfly 44, which is consistent with the failed $L_*$ scenario.





\section{SUMMARY and CONCLUSION}

Analysing deep HST F814W and F105W  images in the HFF, we discovered a large population of the UDGs in two massive galaxy clusters, Abell S1063 and Abell 2744. 
We adopted the UDG selection criteria consistent with those used for nearby galaxy clusters: 
$R_{\rm e,c}>1.5$ kpc, 
\absmue $>23.8$ \SBunit, 
$q>0.3$, 
 and (F814W--F105W) colors bluer than the red boundary of the red sequence.
For comparison, we also selected LSB dwarfs with 1 kpc $<R_{\rm e,c}<1.5$ kpc and the same surface brightness range as the UDGs. 
Primary results are summarized as follows.

\begin{enumerate}
\item We find 
 47 and 40 UDGs in the HST fields of Abell S1063 and Abell 2744, respectively. 
\item The UDGs are mostly located at the faint end of the red sequence in the CMD, showing that they are mostly passively evolving old galaxies.
\item The radial surface brightness profiles of most UDGs are fit well by an exponential law profile: the mean values of the measured S\'{e}rsic indices are  $\langle n\rangle=1.14\pm0.63$ for Abell S1063 and
 $\langle n\rangle=0.94\pm0.65$ for Abell 2744. 
 These are similar to those of Coma UDGs, $\langle n\rangle=1.01\pm0.37$.
\item  The mean values of the elongation parameters of the UDGs are $\langle q\rangle=0.66\pm0.16$ for Abell S1063 and
 $\langle q\rangle=0.68\pm0.17$ for Abell 2744, 
 which are similar to those of Coma UDGs, $\langle q\rangle=0.74\pm0.15$.
\item From the comparison with simple stellar population models, we estimate the stellar mass of the UDGs to range from $10^8$ to $10^9 M_\sun$. 

\item
The radial number density profiles of the UDGs and LSB dwarfs in both clusters show a drop or a flattening in the central region at $r<100$ kpc, while that of the bright galaxies keeps increasing in the central region. 
%
\item
We estimate the enclosed masses within effective radius of the UDGs using the galaxy scaling relations, finding that  
the enclosed masses of most UDGs range from $M(<R_{1/2} ) = 6 \times 10^8 M_\sun$~ to $3 \times 10^9 M_\sun$, and three largest UDGs have much higher masses, $6 \times 10^9 M_\sun$~ to $10^{10} M_\sun$.
From this we find a majority of the UDGs have total masses, $M_{200} = 10^{10}$ to $10^{11}~M_\sun$, and only a few of them
have total masses, $M_{200} = 10^{11}$ to $10^{12}~M_\sun$.   
\item The total number of UDGs within the virial radius of each cluster 
is estimated to be
 N(UDG)$=770\pm114$ 
for Abell S1063, and   
N(UDG)$=814\pm122$ 
for Abell 2744.
Combining these results with data in the literature, we fit the relation between
the total numbers of UDGs and the masses of their host systems with a power law,
N(UDG) $= M_{200}^{0.89\pm0.07}$. 
However, if we fit the data for massive systems with $M_{200}>10^{13} M_\sun$, we obtain N(UDG) $= M_{200}^{1.05\pm0.09}$. 
This value of the power law index is close to one, implying that the efficiency of UDGs
depends little on the mass of their host systems.
%
\item
We conclude on the origin of the UDGs:
A majority of the UDGs 
in Abell S1063 and Abell 2744 have relatively massive dwarf galaxies. 
Only a small number of the UDGs 
 can be massive enough to be failed $L_{*}$ galaxies.
\end{enumerate}
This work was supported by the National Research Foundation of Korea (NRF) grant
funded by the Korean Government (MSIP) (No. 2012R1A4A1028713).
J.K. was supported by the Global Ph.D. Fellowship Program (NRF-2016H1A2A1907015). 
%


\clearpage


\begin{turnpage}
\begin{deluxetable*}{ccccccccccc}
	\tabletypesize{\scriptsize}
	\setlength{\tabcolsep}{0.05in}
	\tablecaption{A Catalog of UDGs in Abell S1063}
	\tablewidth{630pt}
	\tablehead{ \colhead{ID} & \colhead{R.A.} & \colhead{Dec.} & \colhead{\Reff\tablenotemark{a}} & \colhead{$\mu_{\rm e,F814W}$} & \colhead{$F814W$} & \colhead{$F814W-F105W$} & \colhead{n} & \colhead{b/a} & \colhead{\Reffc\tablenotemark{b}} & \colhead{\absmue\tablenotemark{c}} \\
			\colhead{} & \colhead{(J2000)} & \colhead{(J2000)} & \colhead{[kpc]} &\colhead{[\SBunit]} &\colhead{[mag]} &\colhead{[mag]} &\colhead{} &\colhead{} &\colhead{[kpc]} &\colhead{[\SBunit]} }
	\startdata
	AS1063\_UDG001&342.16211&-44.54686& $2.29\pm0.09$ & $25.65\pm0.06$ & $25.61\pm0.02$ & $0.02\pm0.02$ & $0.68\pm0.07$ & $0.51\pm0.02$ & $1.63\pm0.07$ & $24.47\pm0.07$\\
	AS1063\_UDG002&342.16321&-44.54173& $1.61\pm0.14$ & $26.28\pm0.12$ & $26.05\pm0.03$ & $0.11\pm0.02$ & $1.14\pm0.21$ & $0.87\pm0.06$ & $1.50\pm0.14$ & $24.87\pm0.15$\\
	AS1063\_UDG003&342.16568&-44.52137& $2.31\pm0.16$ & $25.85\pm0.14$ & $25.29\pm0.02$ & $0.31\pm0.01$ & $1.84\pm0.21$ & $0.59\pm0.02$ & $1.77\pm0.13$ & $24.21\pm0.15$\\
	AS1063\_UDG004&342.16678&-44.52966& $1.83\pm0.21$ & $25.58\pm0.32$ & $25.31\pm0.01$ & $0.24\pm0.01$ & $1.59\pm0.33$ & $0.68\pm0.07$ & $1.50\pm0.19$ & $24.01\pm0.34$\\
	AS1063\_UDG005&342.16861&-44.51286& $3.73\pm0.28$ & $25.91\pm0.15$ & $24.12\pm0.01$ & $0.31\pm0.01$ & $1.74\pm0.15$ & $0.82\pm0.02$ & $3.38\pm0.26$ & $24.29\pm0.16$\\
	AS1063\_UDG006&342.16876&-44.52851& $2.41\pm0.15$ & $25.97\pm0.08$ & $26.10\pm0.03$ & $0.18\pm0.02$ & $0.83\pm0.14$ & $0.40\pm0.02$ & $1.53\pm0.10$ & $24.71\pm0.11$\\
	AS1063\_UDG007&342.17020&-44.54548& $2.13\pm0.08$ & $25.51\pm0.06$ & $25.24\pm0.02$ & $0.06\pm0.01$ & $0.66\pm0.07$ & $0.62\pm0.02$ & $1.68\pm0.15$ & $24.34\pm0.13$\\
	AS1063\_UDG008&342.17471&-44.52490& $2.41\pm0.80$ & $27.46\pm0.64$ & $26.78\pm0.05$ & $0.18\pm0.03$ & $2.53\pm0.82$ & $0.76\pm0.08$ & $2.10\pm0.07$ & $25.66\pm0.07$\\
	AS1063\_UDG009&342.17581&-44.54522& $2.10\pm0.32$ & $26.45\pm0.32$ & $26.15\pm0.02$ & $0.16\pm0.02$ & $2.55\pm0.45$ & $0.53\pm0.04$ & $1.53\pm0.70$ & $24.64\pm0.66$\\
	AS1063\_UDG010&342.18430&-44.51709& $1.82\pm0.12$ & $25.99\pm0.09$ & $26.00\pm0.03$ & $0.16\pm0.02$ & $0.90\pm0.15$ & $0.70\pm0.04$ & $1.52\pm0.24$ & $24.69\pm0.33$\\
	\enddata
	\tablenotetext{a}{Assuming a distance modulus of $(m-M)_0=41.25$. }
	\tablenotetext{b}{$R_{e,c} = R_e \sqrt{b/a}$ }
	\tablenotetext{c}{\absmue = $\langle\mu\rangle_{\rm e,z} (r') - 10 {\rm log} (1+z) - E(z) - K(z)$ assuming a redshift $z=0.348$.
		\\(This table is available in its entirety in machine-readable form.)}
	\label{tab_1}
\end{deluxetable*}
\end{turnpage}


\begin{turnpage}
\begin{deluxetable*}{ccccccccccc}
	\tabletypesize{\scriptsize}
	\setlength{\tabcolsep}{0.05in}
	\tablecaption{A Catalog of LSB dwarfs in Abell S1063}
	\tablewidth{630pt}
	\tablehead{ \colhead{ID} & \colhead{R.A.} & \colhead{Dec.} & \colhead{\Reff\tablenotemark{a}} & \colhead{$\mu_{\rm e,F814W}$} & \colhead{$F814W$} & \colhead{$F814W-F105W$} & \colhead{n} & \colhead{b/a} & \colhead{\Reffc\tablenotemark{b}} & \colhead{\absmue\tablenotemark{c}} \\
		\colhead{} & \colhead{(J2000)} & \colhead{(J2000)} & \colhead{[kpc]} &\colhead{[\SBunit]} &\colhead{[mag]} &\colhead{[mag]} &\colhead{} &\colhead{} &\colhead{[kpc]} &\colhead{[\SBunit]} }
	\startdata
	AS1063\_LSBdw001&342.15747&-44.54634& $1.70\pm0.06$ & $25.48\pm0.07$ & $26.10\pm0.03$ & $0.48\pm0.02$ & $0.63\pm0.07$ & $0.58\pm0.02$ & $1.29\pm0.05$ & $24.33\pm0.08$\\
	AS1063\_LSBdw002&342.15967&-44.53279& $1.97\pm0.06$ & $25.10\pm0.08$ & $25.61\pm0.02$ & $0.36\pm0.01$ & $0.41\pm0.04$ & $0.46\pm0.02$ & $1.33\pm0.05$ & $24.11\pm0.09$\\
	AS1063\_LSBdw003&342.16046&-44.54249& $1.44\pm0.14$ & $26.27\pm0.11$ & $26.81\pm0.04$ & $0.26\pm0.03$ & $0.93\pm0.22$ & $0.81\pm0.06$ & $1.30\pm0.13$ & $24.95\pm0.15$\\
	AS1063\_LSBdw004&342.16068&-44.53046& $1.67\pm0.13$ & $25.89\pm0.12$ & $25.81\pm0.02$ & $0.05\pm0.02$ & $1.33\pm0.21$ & $0.75\pm0.04$ & $1.45\pm0.12$ & $24.41\pm0.14$\\
	AS1063\_LSBdw005&342.16150&-44.53435& $1.50\pm0.08$ & $25.18\pm0.07$ & $25.55\pm0.02$ & $0.25\pm0.01$ & $0.81\pm0.11$ & $0.62\pm0.03$ & $1.18\pm0.07$ & $23.93\pm0.09$\\
	AS1063\_LSBdw006&342.16214&-44.54063& $1.39\pm0.14$ & $25.79\pm0.13$ & $26.51\pm0.03$ & $0.12\pm0.02$ & $1.26\pm0.29$ & $0.55\pm0.05$ & $1.03\pm0.11$ & $24.33\pm0.17$\\
	AS1063\_LSBdw007&342.16437&-44.54641& $1.51\pm0.11$ & $25.73\pm0.11$ & $26.69\pm0.04$ & $0.16\pm0.03$ & $0.58\pm0.14$ & $0.48\pm0.04$ & $1.05\pm0.09$ & $24.62\pm0.15$\\
	AS1063\_LSBdw008&342.16486&-44.51730& $1.62\pm0.12$ & $25.84\pm0.13$ & $25.84\pm0.02$ & $0.19\pm0.02$ & $1.36\pm0.23$ & $0.79\pm0.05$ & $1.44\pm0.12$ & $24.35\pm0.15$\\
	AS1063\_LSBdw009&342.16562&-44.51790& $1.33\pm0.12$ & $25.92\pm0.10$ & $26.45\pm0.03$ & $0.19\pm0.03$ & $0.70\pm0.18$ & $0.80\pm0.08$ & $1.18\pm0.12$ & $24.72\pm0.15$\\
	AS1063\_LSBdw010&342.16583&-44.53659& $1.21\pm0.07$ & $25.26\pm0.07$ & $25.87\pm0.02$ & $0.21\pm0.01$ & $0.81\pm0.11$ & $0.95\pm0.05$ & $1.18\pm0.07$ & $24.01\pm0.09$\\
	\enddata
	\tablenotetext{a}{Assuming a distance modulus of $(m-M)_0=41.25$. }
	\tablenotetext{b}{$R_{e,c} = R_e \sqrt{b/a}$ }
	\tablenotetext{c}{\absmue = $\langle\mu\rangle_{\rm e,z} (r') - 10 {\rm log} (1+z) - E(z) - K(z)$ assuming a redshift $z=0.348$.
		\\(This table is available in its entirety in machine-readable form.)}
	\label{tab_2}
\end{deluxetable*}
\end{turnpage}


\begin{turnpage}
\begin{deluxetable*}{ccccccccccc}
	\tabletypesize{\scriptsize}
	\setlength{\tabcolsep}{0.05in}
	\tablecaption{A Catalog of UDGs in Abell 2744}
	\tablewidth{630pt}
	\tablehead{ \colhead{ID} & \colhead{R.A.} & \colhead{Dec.} & \colhead{\Reff\tablenotemark{a}} & \colhead{$\mu_{\rm e,F814W}$} & \colhead{$F814W$} & \colhead{$F814W-F105W$} & \colhead{n} & \colhead{b/a} & \colhead{\Reffc\tablenotemark{b}} & \colhead{\absmue\tablenotemark{c}} \\
		\colhead{} & \colhead{(J2000)} & \colhead{(J2000)} & \colhead{[kpc]} &\colhead{[\SBunit]} &\colhead{[mag]} &\colhead{[mag]} &\colhead{} &\colhead{} &\colhead{[kpc]} &\colhead{[\SBunit]} }
	\startdata
	A2744\_UDG001&  3.46229&-30.36563& $2.70\pm0.08$ & $25.70\pm0.06$ & $25.37\pm0.02$ & $0.12\pm0.01$ & $0.20\pm0.03$ & $0.44\pm0.01$ & $1.79\pm0.05$ & $24.81\pm0.14$\\
	A2744\_UDG002&  3.46564&-30.38844& $2.36\pm0.13$ & $26.20\pm0.07$ & $25.51\pm0.02$ & $0.17\pm0.02$ & $0.75\pm0.10$ & $0.63\pm0.03$ & $1.87\pm0.11$ & $25.07\pm0.09$\\
	A2744\_UDG003&  3.46573&-30.38783& $1.99\pm0.08$ & $25.55\pm0.05$ & $24.77\pm0.01$ & $0.28\pm0.01$ & $0.77\pm0.07$ & $0.92\pm0.04$ & $1.91\pm0.09$ & $24.41\pm0.06$\\
	A2744\_UDG004&  3.46982&-30.38590& $4.07\pm0.18$ & $25.68\pm0.08$ & $23.83\pm0.00$ & $0.20\pm0.00$ & $1.46\pm0.08$ & $0.78\pm0.01$ & $3.59\pm0.16$ & $24.24\pm0.08$\\
	A2744\_UDG005&  3.47075&-30.39297& $2.20\pm0.07$ & $25.05\pm0.04$ & $24.74\pm0.01$ & $0.31\pm0.01$ & $0.69\pm0.05$ & $0.52\pm0.01$ & $1.58\pm0.05$ & $23.96\pm0.05$\\
	A2744\_UDG006&  3.47095&-30.36673& $2.71\pm0.31$ & $27.01\pm0.21$ & $26.96\pm0.04$ & $0.10\pm0.03$ & $1.47\pm0.29$ & $0.42\pm0.03$ & $1.76\pm0.21$ & $25.57\pm0.23$\\
	A2744\_UDG007&  3.47105&-30.37178& $2.07\pm0.06$ & $25.55\pm0.07$ & $25.57\pm0.02$ & $0.23\pm0.01$ & $0.48\pm0.04$ & $0.56\pm0.02$ & $1.55\pm0.05$ & $24.60\pm0.08$\\
	A2744\_UDG008&  3.47224&-30.38179& $1.79\pm0.09$ & $25.90\pm0.08$ & $26.01\pm0.03$ & $0.13\pm0.02$ & $0.67\pm0.09$ & $0.83\pm0.04$ & $1.63\pm0.09$ & $24.82\pm0.10$\\
	A2744\_UDG009&  3.47296&-30.37895& $2.20\pm0.11$ & $25.27\pm0.07$ & $24.95\pm0.01$ & $0.04\pm0.01$ & $1.19\pm0.11$ & $0.55\pm0.02$ & $1.63\pm0.09$ & $23.93\pm0.08$\\
	A2744\_UDG010&  3.48256&-30.39862& $3.17\pm0.13$ & $26.42\pm0.06$ & $25.66\pm0.03$ & $0.16\pm0.02$ & $0.76\pm0.09$ & $0.50\pm0.02$ & $2.24\pm0.10$ & $25.29\pm0.08$\\	
	\enddata
	\tablenotetext{a}{Assuming a distance modulus of $(m-M)_0=40.94$. }
	\tablenotetext{b}{$R_{e,c} = R_e \sqrt{b/a}$ }
	\tablenotetext{c}{\absmue = $\langle\mu\rangle_{\rm e,z} (r') - 10 {\rm log} (1+z) - E(z) - K(z)$ assuming a redshift $z=0.308$.
		\\(This table is available in its entirety in machine-readable form.)}
	\label{tab_3}
\end{deluxetable*}
\end{turnpage}


\begin{turnpage}
\begin{deluxetable*}{ccccccccccc}
	\tabletypesize{\scriptsize}
	\setlength{\tabcolsep}{0.05in}
	\tablecaption{A Catalog of LSB dwarfs in Abell 2744}
	\tablewidth{630pt}
	\tablehead{ \colhead{ID} & \colhead{R.A.} & \colhead{Dec.} & \colhead{\Reff\tablenotemark{a}} & \colhead{$\mu_{\rm e,F814W}$} & \colhead{$F814W$} & \colhead{$F814W-F105W$} & \colhead{n} & \colhead{b/a} & \colhead{\Reffc\tablenotemark{b}} & \colhead{\absmue\tablenotemark{c}} \\
		\colhead{} & \colhead{(J2000)} & \colhead{(J2000)} & \colhead{[kpc]} &\colhead{[\SBunit]} &\colhead{[mag]} &\colhead{[mag]} &\colhead{} &\colhead{} &\colhead{[kpc]} &\colhead{[\SBunit]} }
	\startdata
	A2744\_LSBdw001&  3.45478&-30.37182& $1.59\pm0.06$ & $25.22\pm0.07$ & $25.73\pm0.02$ & $0.40\pm0.01$ & $0.45\pm0.04$ & $0.83\pm0.03$ & $1.45\pm0.37$ & $24.29\pm0.63$\\
	A2744\_LSBdw002&  3.45811&-30.36477& $1.39\pm0.10$ & $25.66\pm0.09$ & $26.27\pm0.02$ & $0.07\pm0.02$ & $0.94\pm0.19$ & $0.54\pm0.04$ & $1.02\pm0.06$ & $24.43\pm0.08$\\
	A2744\_LSBdw003&  3.45917&-30.38811& $1.41\pm0.08$ & $26.41\pm0.13$ & $27.28\pm0.06$ & $0.26\pm0.04$ & $0.29\pm0.08$ & $0.88\pm0.06$ & $1.32\pm0.09$ & $25.59\pm0.13$\\
	A2744\_LSBdw004&  3.45989&-30.37544& $1.21\pm0.06$ & $25.01\pm0.07$ & $25.41\pm0.02$ & $0.15\pm0.01$ & $0.75\pm0.10$ & $0.77\pm0.03$ & $1.07\pm0.09$ & $23.88\pm0.13$\\
	A2744\_LSBdw005&  3.46269&-30.37417& $1.21\pm0.06$ & $25.51\pm0.10$ & $26.46\pm0.03$ & $0.25\pm0.02$ & $0.51\pm0.07$ & $0.80\pm0.04$ & $1.08\pm0.06$ & $24.54\pm0.09$\\
	A2744\_LSBdw006&  3.46637&-30.39392& $1.63\pm0.09$ & $25.43\pm0.10$ & $26.20\pm0.03$ & $0.08\pm0.02$ & $0.30\pm0.06$ & $0.48\pm0.02$ & $1.13\pm0.06$ & $24.61\pm0.11$\\
	A2744\_LSBdw007&  3.46743&-30.36893& $1.39\pm0.09$ & $25.51\pm0.07$ & $25.69\pm0.02$ & $0.36\pm0.01$ & $1.00\pm0.14$ & $0.82\pm0.04$ & $1.26\pm0.06$ & $24.25\pm0.10$\\
	A2744\_LSBdw008&  3.47124&-30.37027& $1.28\pm0.08$ & $25.26\pm0.08$ & $25.59\pm0.02$ & $0.18\pm0.01$ & $1.20\pm0.17$ & $0.71\pm0.04$ & $1.08\pm0.08$ & $23.92\pm0.10$\\
	A2744\_LSBdw009&  3.47139&-30.36647& $1.76\pm0.15$ & $25.74\pm0.14$ & $25.86\pm0.02$ & $0.04\pm0.01$ & $1.80\pm0.23$ & $0.47\pm0.03$ & $1.21\pm0.07$ & $24.20\pm0.10$\\
	A2744\_LSBdw010&  3.47548&-30.36458& $1.33\pm0.14$ & $26.39\pm0.12$ & $27.28\pm0.06$ & $0.24\pm0.04$ & $0.83\pm0.24$ & $0.61\pm0.05$ & $1.04\pm0.11$ & $25.22\pm0.15$\\ 
	\enddata
	\tablenotetext{a}{Assuming a distance modulus of $(m-M)_0=40.94$. }
	\tablenotetext{b}{$R_{e,c} = R_e \sqrt{b/a}$ }
	\tablenotetext{c}{\absmue = $\langle\mu\rangle_{\rm e,z} (r') - 10 {\rm log} (1+z) - E(z) - K(z)$ assuming a redshift $z=0.308$.
		\\(This table is available in its entirety in machine-readable form.)}
	\label{tab_4}
\end{deluxetable*}
\end{turnpage}

\end{document}